# Cost-efficiency trade-offs of the human brain network revealed by a multiobjective evolutionary algorithm

Junji Ma[1], Jinbo Zhang[1], Ying Lin[1*], Zhengjia Dai[1*]

[1] Department of Psychology, Sun Yat-sen University, Guangzhou, China, 510006

*Corresponding Author:

Zhengjia Dai, Ph.D., Department of Psychology, Sun Yat-sen University, Guangzhou, China, 510006. Email: daizhengj@mail.sysu.edu.cn

Ying Lin, Ph.D., Department of Psychology, Sun Yat-sen University, Guangzhou, China, 510006. Email: linying23@mail.sysu.edu.cn

**Abstract**

It is widely believed that the formation of brain network structure is under the pressure of optimal trade-off between reducing wiring cost and promoting communication efficiency. However, the question of whether this trade-off exists in empirical human brain networks and, if so, how it takes effect is still not well understood. Here, we employed a multiobjective evolutionary algorithm to directly and quantitatively explore the cost-efficiency trade-off in human brain networks. Using this algorithm, we generated a population of synthetic networks with optimal but diverse cost-efficiency trade-offs. It was found that these synthetic networks could not only reproduce a large portion of connections in the empirical brain networks but also embed a resembling small-world structure. Moreover, the synthetic and empirical brain networks were found similar in terms of the spatial arrangement of hub regions and the modular structure, which are two important topological features widely assumed to be outcomes of cost-efficiency trade-offs. The synthetic networks had high robustness against random attack as the empirical brain networks did. Additionally, we also revealed some differences of the synthetic networks from the empirical brain networks, including lower segregated processing capacity and weaker robustness against targeted attack. These findings provide direct and quantitative evidence that the structure of human brain networks is indeed largely influenced by optimal cost-efficiency trade-offs. We also suggest that some additional factors (e.g., segregated processing capacity) might jointly determine the network organization with cost and efficiency.

**Introduction**

The brain is a complex system that can be characterized as a network with numerous connections linking brain regions, namely the brain connectome (Bullmore and Sporns, 2009, 2012; Sporns et al., 2005). Graph theoretical approaches in network neuroscience have revealed several pronounced topological features of the brain connectome, including small-worldness (for review see Bassett and Bullmore, 2006; Bassett and Bullmore, 2016; Liao et al., 2017), modular organization (Meunier et al., 2010; Meunier et al., 2009) and the presence of hub regions (for review see van den Heuvel and Sporns, 2013). Although these features do provide a systematic description of brain network architecture, the underlying mechanism that shapes the brain structure remains poorly understood.

Wiring cost has long been recognized as one fundamental factor that shapes the brain network structure. Embedding within the limited space of the skull, construction and maintenance of anatomical connections in brain networks consume a high material cost (Ahn et al., 2006; Mitchison, 1991). Considering the cost constraint, previous studies have pointed out that several features of brain networks might be the outcome of minimizing cost. First, it was commonly found that close region pairs had a much higher probability of connecting with each other than remote region pairs that require costly long connections to be interconnected (Klyachko and Stevens, 2003; Markov et al., 2013; Young, 1992). Moreover, the existence of some topological features (e.g., high modularity and hierarchical organization) was also attributable to the pressure of minimizing cost (Samu et al., 2014). However, in silico, the wiring cost of empirical brain networks was found to be nearly minimal but not strictly minimized compared with simulated networks (Kaiser and Hilgetag, 2006). Hence, minimizing cost is one, but not the only, fundamental factor shaping brain network organization.

Efficient information processing and communication are also required for brain networks (Laughlin and Sejnowski, 2003). The small-worldness of brain networks enables them to achieve an optimal balance between local processing (i.e., high clustering coefficient) and global communication (i.e., short characteristic path length) efficiency (for review, see Bassett and Bullmore, 2006; Bassett and Bullmore, 2016; Liao et al., 2017). Moreover, the existence of long-distance connections (Kaiser and Hilgetag, 2006) that promote communication between remote regions, modular structure (Meunier et al., 2010; Meunier et al., 2009) and highly connected hubs (van den Heuvel and Sporns, 2013) that promote segregated and integrated information processing, all reflect an optimal arrangement for efficient network communication. However, these features inevitably increase the cost of the brain network. Therefore, the network structure configuration does not simply minimize cost or promote efficiency but seeks the optimal trade-off between these two factors (Bullmore and Sporns, 2012).

Recent studies have attempted to examine the hypothesis of optimal cost-efficiency trade-off in non-human brain networks (Chen et al., 2013; Chen et al., 2017). They defined an optimization objective that described the cost-efficiency trade-off as a weighted sum of the cost and efficiency measures, and then generated synthetic brain networks by employing the simulated annealing algorithm to approximate the optimal solution under the above objective. The comparison between the synthetic networks and the empirical macaque/C. elegans brain network found that the synthetic networks could reproduce the modular organization and the spatial location of several hubs in the empirical network. Such findings strongly supported the hypothesis of optimal cost-efficiency trade-off in non-human brain networks. However, whether optimizing the cost-efficiency trade-off is also the underlying force that drives the formation of human brain networks remains an open question. Moreover, using fixed weights to quantify the relative importance of cost and efficiency, as did in previous research (Chen et al., 2013; Chen et al., 2017), could induce biased or incomprehensive conclusions when discussing the effect of cost-efficiency trade-offs in the formation of brain networks. First, the relative importance of cost and efficiency is actually unknown. Presuming their weights may hinder full exploration of possible cost-efficiency trade-offs in empirical brain networks. Second, using fixed weights to define the cost-efficiency trade-off inherently denies the possibility of variation in the trade-off. However, the diversity of brain network structure is an adaptive property from an evolutionary perspective, which can prevent the risk of extinction in species while confronting environmental changes (Frankel et al., 2014; Schindler et al., 2010). Hence, the variation in the corresponding cost-efficiency trade-off may also exist and is worth further consideration.

In the current study, we explored the cost-efficiency trade-off in the human brain network using a novel approach that can avoid explicitly defining the relative importance between cost and efficiency. First, we modeled the cost-efficiency trade-off as a multiobjective optimization problem aimed at minimizing the wiring cost while simultaneously maximizing communication efficiency. Then, we employed a multiobjective evolutionary algorithm (MOEA) to solve the above problem, which generated a set of synthetic networks with different but equally good trade-offs (namely, the nondominated set) between the cost and efficiency objectives by imitating the evolutionary process in nature. After using the MOEA to generate a population of synthetic networks with optimal cost-efficiency trade-off, we then examined how the cost-efficiency trade-off influences human brain network organization by evaluating the similarity between the synthetic and empirical networks.

**Materials and Methods**

*Anatomical Brain Network Construction*

Nodes and edges are two basic network elements. In this study, ninety nodes covering the whole brain were defined by Automated Anatomical Labeling parcellation (AAL, Tzourio-Mazoyer et al., 2002). Here, we acquired dMRI images of 93 healthy participants to construct empirical brain networks. The AAL atlas was originally defined in the standard Montreal Neurological Institute (MNI) space and was transformed into the native diffusion space for each participant with the typical procedure of previous studies (Gong et al., 2009). Specifically, individual FA images were coregistered to the T1-weighted images, and the T1-weighted images were normalized to the ICBM152 T1 template in the MNI space. These two transformation matrices were then inverted to warp the AAL atlas from MNI space to native diffusion space for each participant. To define network edges, we performed deterministic tractography based on fiber assignment by continuous tracking (FACT) using the Diffusion Toolkit (http://trackvis.org) (Wang et al., 2007). In the tracking procedure, a seed was distributed at the center of each voxel with an FA value greater than 0.2 in the WM mask. A streamline was started from each seed and terminated if the streamline reached a voxel with a turning angle greater than 45°, the FA value was less than 0.2, or the streamline entered a voxel out of the WM mask. For each pair of nodes, the edge was defined as one if there was at least one streamline between them; otherwise, it was set to zero. Therefore, for each participant, we obtained one 90 by 90 binary structural network.

*Objective Functions of Brain Network*

In this study, to construct synthetic networks with the optimal cost-efficiency trade-offs, we first defined two objective functions to evaluate the wiring cost and the processing efficiency of brain networks. The first objective $F_c$, which measured the total wiring cost of the brain network, was calculated as follows:

$$F_c(A) = \frac{1}{2}\sum_{i=1}^{N}\sum_{j=1}^{N} d_{ij}a_{ij}, i \neq j = 1,2,3......N$$

where $d_{ij}$ is the Euclidean distance between the centroids of nodes $i$ and $j$; $A = [a_{ij}]_{N \times N}$ is the adjacency matrix of the binary structural network and $a_{ij}$ is the indicator value for the existence of the edge between nodes $i$ and $j$, where 1-value indicates edge exist between nodes and 0-value indicates no edge; $N$ is the total number of nodes in the network (here, $N = 90$). To be noted, we used the Euclidean distance between two regions to approximate the wiring cost of their connection.

The second objective measured the processing efficiency of brain networks. Notably, to unify the optimization direction with the cost objective (i.e., minimization), we used the path length of the synthetic network to define the efficiency objective $F_e$. The smaller the $F_e$ value is, the higher efficiency the network possesses. The detailed calculation of $F_e$ is shown below:

$$F_e(A) = \frac{1}{2}\sum_{i=1}^{N}\sum_{j=1}^{N} l_{ij}, i \neq j = 1,2,3......N$$

where $l_{ij}$ is the shortest path length between nodes $i$ and $j$ according to the binary network adjacency matrix $A$ and $N$ is the total number of nodes in the network (here, $N = 90$).

*Cost-efficiency Balanced Network Construction*

After defining the two objective functions as above, we adopted the nondominated sorting genetic algorithm (NSGA-II; Deb et al., 2002), a popular multiobjective evolutionary algorithm (MOEA), to construct synthetic networks with the optimal cost-efficiency trade-offs. Without predefining the weight of relative importance between cost and efficiency, MOEA evolves a population of candidate networks to seek the best trade-off between these two objectives. The overall procedure of NSGA-II we implemented is as follows: (1) randomly initialize the connection matrix of each synthetic network (2) exchange edges from high quality synthetic networks from last generation (3) mutated the synthetic networks with a low probability (4) select the networks with dominated performance in cost-efficiency trade-off for next generation (5) terminate the iteration if synthetic network stop evolving.

Since MOEA is a probabilistic algorithm, the results can vary across different runs. We ran the algorithm 30 independent times and then used the fast nondominated sorting approach to select the nondominated solutions among the results of all the runs for subsequent analyses.

*Recovery Rate of Synthetic Networks*

The above NSGA-II generated 263 synthetic networks that approximated the optimal trade-off between cost and efficiency. To measure the similarity between the synthetic networks and the empirical human brain networks, we computed the recovery rate $R$ of the synthetic networks by calculating the ratio of overlapping entries between synthetic and empirical adjacency matrices (Chen et al., 2013; Costa Lda et al., 2007). The recovery rate was computed as follows:

$$R = \sqrt{R_0 R_1} \; ; \; R_0 = \frac{K_{r0}}{K_0} \, , R_1 = \frac{K_{r1}}{K_1}$$

where $R_0$ and $R_1$ are the recovery rates of zero- and one-entries of the matrix, respectively; $K_0$ and $K_1$ are the numbers of zero- and one-entries in the matrix of the empirical brain network (without considering the diagonal entries); $K_{r0}$ and $K_{r1}$ are the numbers of overlapping zero- and one-entries between the synthetic and empirical networks.

According to previous studies (Chen et al., 2013; Chen et al., 2017), edges with different physical distances or from different modules might play different roles in the cost-efficiency trade-off. We further calculated the recovery rate on different subgroups of edges: distance subgroups according to the Euclidean distance between region pairs and modular subgroups according to the network partition obtained by the method detailed in the "*Modular Structure*" section.

*Topological Characteristics of the Synthetic Networks*

To further examine whether the synthetic networks could capture the topological features of empirical brain networks, we calculated the graphical metrics at the whole-brain level, including the small-world metrics [i.e., clustering coefficient (Cp), characteristic path length (Lp), normalized clustering coefficient ($\gamma$), normalized characteristic path length ($\lambda$), and small-worldness ($\sigma$)], efficiency metrics [i.e., global efficiency (Eg) and local efficiency (El)] and modularity metrics [i.e., modularity (Q) and number of modules (Mn)]. A high clustering coefficient, normalized clustering coefficient, local efficiency, modularity and number of modules reflect a better capacity of segregated processing in the network. Low characteristic path length, normalized clustering coefficient and high global efficiency characterize a better capacity of integrated processing in the network. High small-worldness shows a better balance between segregated and integrated processing in the brain network. Calculation of the above graphical metrics was performed using the Graph Theoretical Network Analysis toolbox (GRETNA; Wang et al., 2015) and Brain Connectivity Toolbox (BCT; Rubinov and Sporns, 2010). Notably, considering the potential effect of cost on these metrics, only the synthetic networks whose cost objective values were distributed in the cost range of the SCNU sample were selected for comparison. All seventy-five synthetic networks within this cost range were finally selected in the current analysis. Moreover, we also select a representative synthetic network from the population for subsequent analyses.

*Degree Centrality and Hubs*

Hub regions, which have a relatively large number of connections, are commonly suggested to have high cost and high communication capacity in human brain networks (for review, see van den Heuvel and Sporns, 2013). Therefore, the existence of hub regions might be the outcome of the cost-efficiency trade-off. To examine this hypothesis, we compared the generation and spatial arrangement of hub regions between synthetic and empirical networks. In the current study, we first examined the similarity in the degree distribution between the representative synthetic and empirical networks. Then, the regions that ranked in the top 20% in descending order of degree centrality (Zuo et al., 2012) were identified as hubs of each representative network, and their spatial locations were compared.

*Modular Structure*

The human brain network is organized into a modular structure with clusters (i.e., modules) of densely connected nodes and sparse connections between these clusters (Meunier et al., 2009). In this section, we investigated whether the synthetic network with optimal cost-efficiency trade-off also possessed a modular structure and how similar the modular partition was with that of the empirical brain network. First, we applied the Louvain community detection algorithm (Blondel et al., 2008) on the representative empirical and synthetic networks to obtain their modular partitions respectively. Then, we computed the mutual information between the two partitions to evaluate the similarity in the modular structure. Furthermore, to examine whether different modules were differently influenced by the cost-efficiency trade-off, we computed the recovery rate on connections within and between modules based on the modular partition of the empirical brain network, respectively. To evaluate the recovery ability under different functional modules, we also computed the modular level recovery rate on a commonly used template that was derived from functional networks (Shen et al., 2013). By assigning each region to the module with the maximum area of overlapping voxels, the 90 regions were assigned to seven functional modules, including the medial frontal network (MFN), frontoparietal network (FPN), default mode network (DMN), subcortical network (subcortical), motor network (motor), visual network (visual), and visual association network (VAN).

*Robustness of Network*

Robustness is an adaptive property of brain networks, which reflects the ability to preserve normal network functions under attacks (e.g., lesions). To investigate whether robustness also existed in the synthetic networks constructed under the optimal trade-offs between cost and efficiency, we performed computational attacks on the representatives of synthetic and empirical brain networks respectively, and compared the resulting network degradation in terms of global efficiency and local efficiency. Two types of computational attacks were conducted: random attacks and targeted attacks. For the random attacks, we randomly deleted nodes step by step with a step of 10% nodes in the entire network. Since node deletion involved certain randomness, at each step, we ran the procedure 100 times and obtained 100 attacked networks. For the targeted attacks, we deleted nodes in descending order of degree centrality step by step, with the step length set the same as that in the random attack.

**Results**

*Global Properties of Synthetic Networks Under Optimal Cost-efficiency Trade-offs*

Using the NSGA-II algorithm, we obtained a set of 263 synthetic networks optimized under the cost-efficiency trade-off. Fig. 1A displays the distribution of these synthetic networks (blue points) and the empirical networks (orange points) in the cost-efficiency morphospace. The 93 empirical brain networks also composed a front along a similar direction but were relatively narrower and were fully dominated by that of the synthetic networks. Together, visual inspection showed that the synthetic networks and empirical brain networks had similar distributions within the cost-efficiency morphospace, but the synthetic networks achieved better and more diverse cost-efficiency trade-offs.

To quantify the similarity between the synthetic networks and the empirical brain networks, we employed the recovery rate to evaluate the ratio of overlapping edges at the whole-brain level. As shown in Fig. 1B, the synthetic networks had a high overall recovery rate of the empirical networks' connections ($R = 0.536 \pm 0.131$). To examine the significance of the recovery rate, we generated a sample of random networks (sample size = 263, same as the number of synthetic networks) that preserved the number of edges and degree distribution of the group-level empirical brain network. The two-sample t-test revealed that the synthetic networks achieved a significantly higher recovery rate than the random networks (t = 19.885, p < 0.001), which suggested a significant similarity with the empirical brain networks.

In addition to the recovery rate of connections, we also examined the topological characteristics of the synthetic networks at the whole-brain level. Note that to avoid the potential effect of cost on these metrics, only the 75 synthetic networks whose cost objective values were in the same range as the empirical data (cost range: 15,958 - 27,544) were selected for analyses. The difference between the cost of the selected synthetic networks and empirical networks was not significant (two-sample test: t = -0.912, p = 0.360). For the small-world related metrics (Fig. 1C), both the synthetic networks and the empirical brain networks were organized as small-world networks (mean $\sigma > 1$), but some differences did exist between the two groups. The two-sample t-test showed that the synthetic networks scored lower in terms of the clustering coefficient (t = -18.688, p < 0.001), characteristic path length (t = -40.379, p < 0.001), normalized clustering coefficient (t = -52.910, p < 0.001) and normalized characteristic path length (t = -70.784, p < 0.001) than the empirical brain networks. Moreover, the small-worldness was also lower in the synthetic networks than in the empirical brain networks (t = -50.942, p < 0.001). For efficiency metrics (Fig. 1D), the synthetic networks showed higher global efficiency (t = 43.354, p < 0.001) but relatively lower local efficiency (t = -11.156, p < 0.001) than the empirical brain networks. For modularity metrics (Fig. 1E), the synthetic networks showed a weaker modular structure (t = -92.839, p < 0.001) and had fewer modules (t = -12.387, p < 0.001) compared to the empirical brain networks.

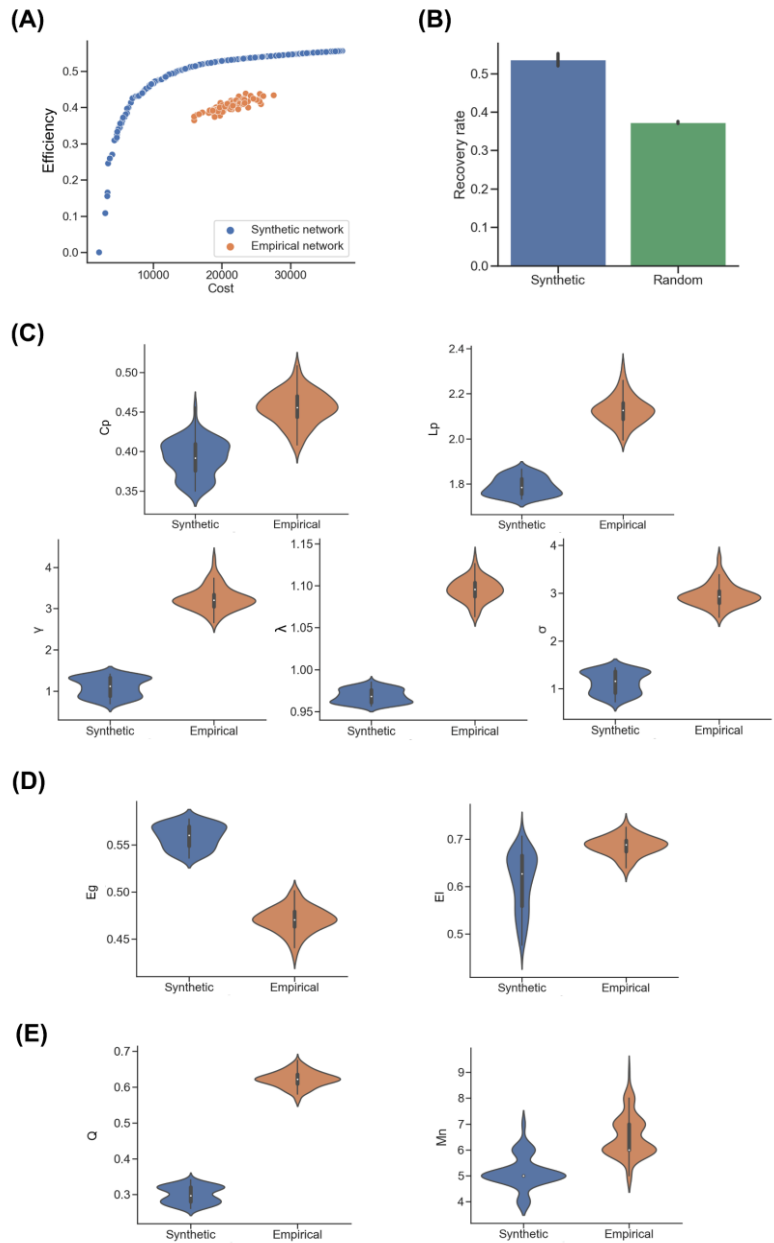

**Figure 1**. Network-level characteristics of cost-efficiency networks. (A) Distribution of synthetic networks (blue points) and empirical brain networks (orange points) in morphospace of cost and efficiency. (B) Recovery rate of synthetic networks and random networks. (C) Clustering coefficient (Cp), characteristic path length (Lp), Normalized clustering coefficient (γ), normalized characteristic path length (λ) and small-worldness (σ) of synthetic networks and empirical brain networks. (D) Global efficiency (Eg) and local efficiency (El) of synthetic networks and empirical brain networks. (E) Modularity (Q) and module number (Mn) of synthetic networks and empirical brain networks.

*Distance-Dependent Connections Pattern of Synthetic Network Under Cost-efficiency Trade-offs*

The above results on the recovery rate revealed a high similarity in the overall connection patterns between the synthetic networks and the empirical brain networks. However, the detailed features of the recovered connections remain unknown. To address this issue, we performed further analyses on the representatives of synthetic networks (density = 13.71%) and empirical brain networks (density = 9.69%). The adjacency matrices and the brain maps of both representative networks are shown in Fig. 2A. A visual inspection found that the patterns were quite similar between the two networks, except that the synthetic network had some obvious global hubs (i.e., regions connected with almost all other regions in the network). The overall recovery rate of the representative synthetic network also supported its similarity with the empirical brain network ($R = 0.638$).

Although the overall recovery rate was high, we found that the recovery of connections was not uniform across the entire network. After dividing the edges into several distance intervals (Fig. 2B), we observed that the representative synthetic network recovered a large proportion of short-distance edges (Euclidian distance <= 40 mm, $0.510 < Rs < 0.860$) but a relatively small proportion of middle- or long-distance edges (Euclidian distance > 40 mm, $Rs < 0.455$). We then divided the entries in the synthetic adjacency matrix into two subgroups based on whether they represented connections (1-entries) or not (0-entries) in the empirical brain network. Similar to the pattern in Fig. 2B, the recovery rate of one-entries ($R1$) was higher in short-distance edges than in the middle- or long-distance edges (Fig. 2C). The recovery rate of zero-entries ($R0$) was high in most distance subgroups, except for the distance range of 20-40 mm (Fig. 2C), indicating a relatively high proportion of false-positive edges within this range. Actually, as shown in Fig. 2D, except for very short edges (distance < 20 mm), the number of edges in the empirical brain network decreased as a function of distance, and only a few long-distance edges existed. Thus, the number of unrecovered edges, especially long-distance edges, was relatively small under the optimization of the cost-efficiency trade-off. In general, the above findings suggested that local connections are preferable while trading off between cost and efficiency, which leads to the recovery of most local connections in the human brain network.

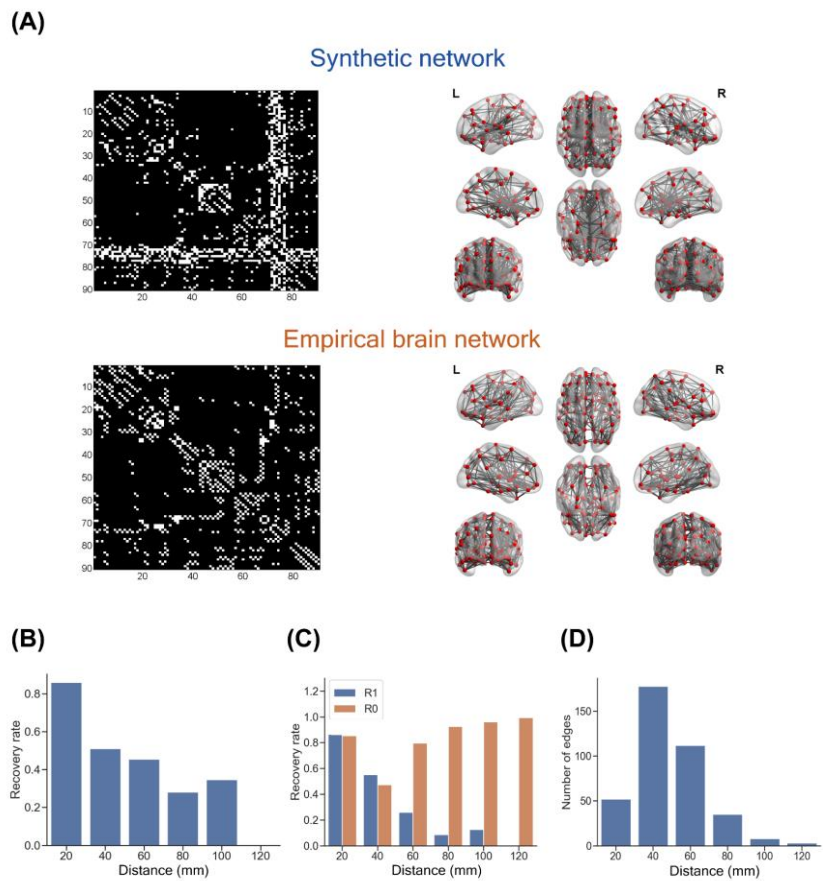

**Figure 2**. Representative synthetic network. (A) Adjacency matrices (left panel) and brain maps (right panel) of the representative synthetic network and the group-level empirical brain network. White color in the adjacency matrices indicates the existence of edges. (B) Overall recovery rate of edges with different distance. (C) Recovery rate of one-entries (R1) and zero-entries (R0) (left panel) (D) number of edges in the group-level empirical brain network.

*Degree Distribution and Hub Regions of Synthetic Network Under Cost-efficiency Trade-offs*

After examining the recovery rates of connections, we further investigated the specific topological characteristics that emerged with the recovered connections. The first one we investigated was the presence of hubs. To identify hub regions, we first computed the degree centrality of the brain regions and sorted them according to their degree centrality. As shown in Fig. 3A, the overall pattern of the degree distribution was similar between the representatives of synthetic networks and empirical brain networks. That is, in both distributions, most regions had relatively low degree centrality, while a small number of regions owned a disproportionally large number of connections. The Kolmogorov-Smirnov test further supported the similarity between these two distributions ($p = 0.675$). Nonetheless, the two degree distributions showed some difference in the scale of degree centrality. After identifying the hub regions (i.e., the top 20% regions with the highest degree centrality) in the two representative networks, Fig. 3B compared their spatial locations. The hub regions largely overlapped in the left precuneus, left olfactory cortex, left insula, left caudate and bilateral putamen (Fig. 3B right panel).

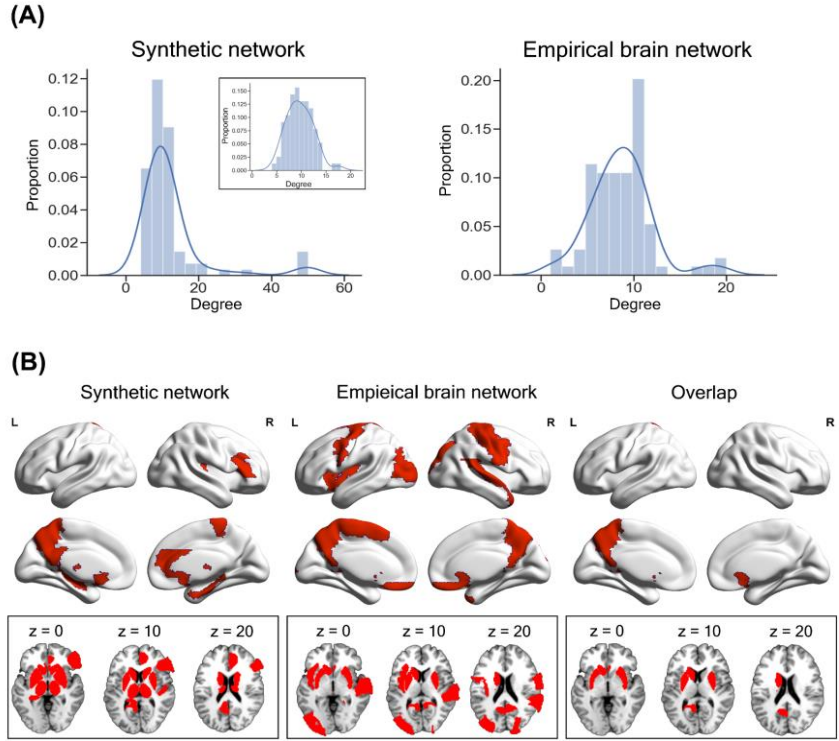

**Figure 3**. Degree and hubs of cost-efficiency network. (A) Degree distribution of the representative synthetic network and empirical brain network. Inset of left panel is the degree distribution within 0~20, same range as the empirical network (B) Spatial distribution of hub regions (red areas) in the synthetic network, empirical brain network and the overlapped hub regions between them.

*Modular Structure of Synthetic Network Under Cost-efficiency Trade-offs*

Another topological feature we considered is the network modular structure, which is commonly suggested to be associated with the cost-efficiency trade-off (Bullmore and Sporns, 2012). Applying the Louvain community detection algorithm, we obtained the modular partitions of the representative synthetic network and the group-level empirical brain network (Fig. 4A). The modular partitions were similar between these two networks, with mutual information at 0.434. Based on the modular partitions of the empirical brain network, we examined the recovery of intra- and inter-module connections in the synthetic network, and found that the recovery rates were not uniform across modules (Fig. 4B). Even in the same module, the recovery rate was quite different between inter- and intra-module connections. Furthermore, Fig. 4C shows that the recovery rate also varied obviously across functional modules defined by Shen et al. (2013). Specifically, intra-module connections had a relatively higher recovery rate (R = 0.691) than inter-module connections (R = 0.605; Fig. 4C right panel). The variety in recovery rate across modules suggested that different modules were affected by cost-efficiency trade-offs to different degrees.

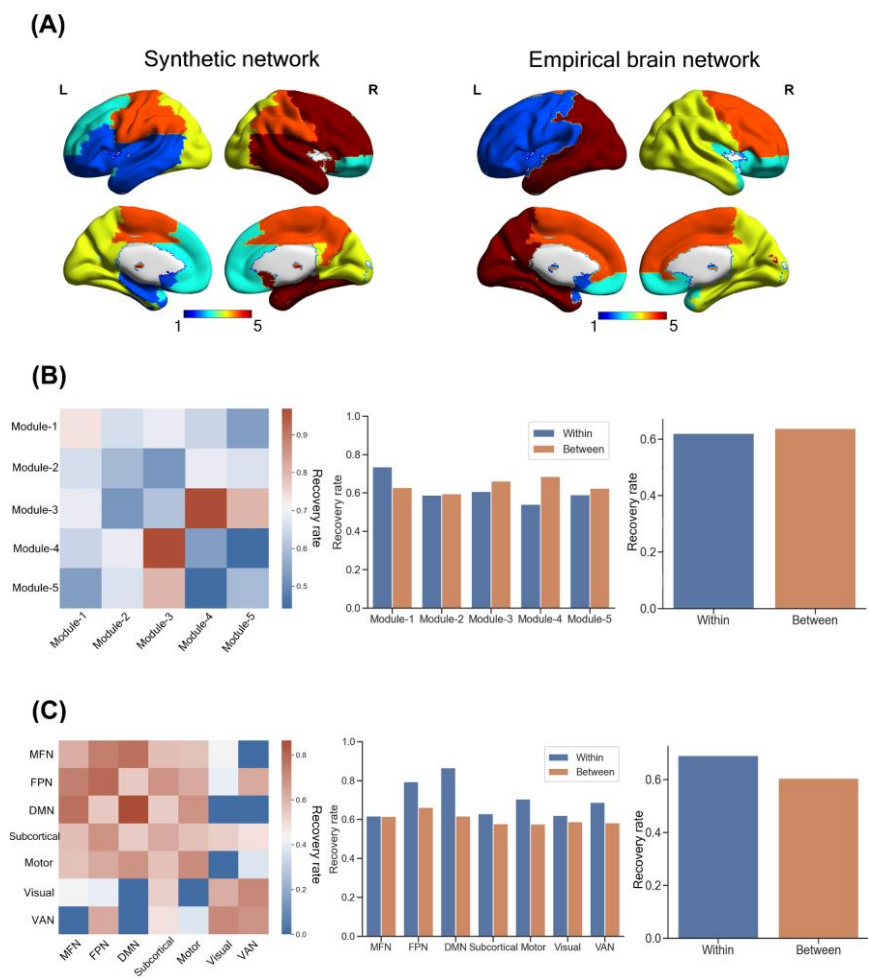

**Figure 4**. Modular structural of cost-efficiency network. (A) Modular partition of the representative synthetic network and empirical brain network. (B) Recovery rate of within- or between-module edges under the modular partition of empirical structural network. (C) Recovery rate of within- or between-module edges under a commonly used functional modular partition (Shen et al., 2013).

*Network Robustness of Synthetic Network Under Cost-efficiency Trade-offs*

Finally, we investigated the robustness, which is an important property of the brain network, of both the representatives of synthetic networks and empirical brain networks. As shown in Fig. 5A, under almost every step of the random attacks, the synthetic network had slightly better robustness than the empirical brain network in terms of both global efficiency (ts > 2.710, ps < 0.007) and local efficiency (ts > 2.411, ps < 0.017) degeneration. The two networks achieved similar degeneration in terms of local efficiency (t = 1.761, p = 0.080) only at the step that removed 80% of the nodes. However, the synthetic network was much more vulnerable to targeted attacks on high-degree regions than the empirical brain network (Fig. 5B), where global and local efficiency degenerated more rapidly. In other words, high-degree regions, especially those global hubs, played a much more important role in the synthetic network communication, leading to higher overall fragility of the network.

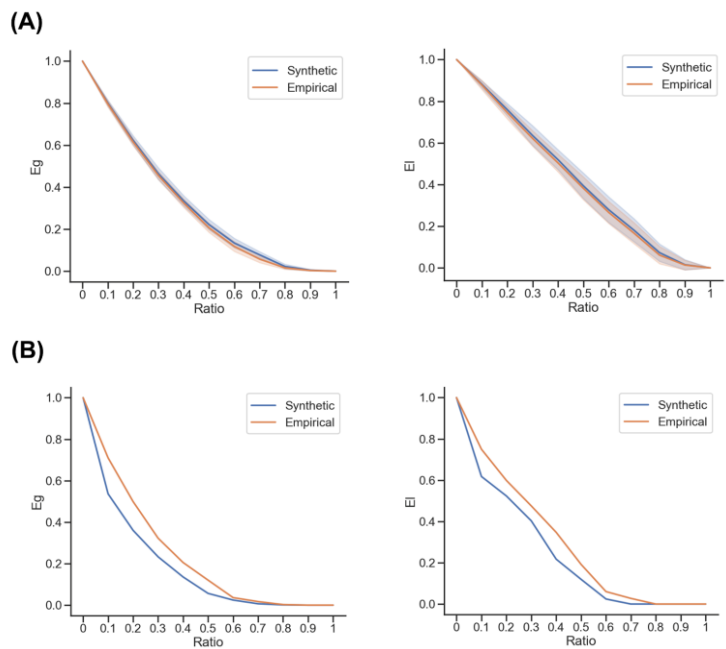

**Figure 5**. Degeneration of networks under computational attack. (A) Degeneration of global efficiency

(Eg) and local efficiency (El) under random attack on the synthetic and empirical brain networks. (B) Degeneration of global efficiency (Eg) and local efficiency (El) under targeted attack on the synthetic and empirical brain network. To be noted, the 'Ratio' indicates ratio of deleted nodes in the current step and shadow of lines indicate range of one standard deviation.

**Discussion**

In this study, we adopted an MOEA approach to reconstruct a population of synthetic networks that simultaneously optimized wiring cost and communication efficiency. These synthetic networks showed good recovery of connection patterns in the empirical brain networks. Through a comparison between the synthetic networks and the empirical brain networks, we revealed a series of important network features that are related to cost-efficiency trade-offs. First, most of the connections in the empirical brain networks, especially the local connections, were recovered in the synthetic networks. Second, similar to the empirical brain networks, the synthetic networks showed a clear small-world property. Third, the modular structure and the spatial distribution of hub regions in the synthetic network were also similar to the empirical brain networks. Fourth, compared to the empirical brain networks, the synthetic networks achieved similar or even better robustness under random attack. The synthetic networks also deviated from the empirical brain networks from a few aspects, including relatively weaker segregated processing capacity (i.e., lower average clustering coefficient and local efficiency) and weaker robustness under targeted attack.

In general, our current study support the notion that cost-efficiency trade-off shapes the structure of human brain network. Further, we reveal several important features (e.g., modular structure) generated under the pressure of cost-efficiency trade-off. Finally, some additional factors, such as robustness, might be additional factors in the trade-off shaping brain network.